\documentclass{article}[12pt]
\usepackage[pdftex]{graphicx}   
\begin{document}
\title{Twirling Operations to Produce Energy Eigenstates of a Hamiltonian by Classically Emulated Quantum Simulation }
\author{Kazuto Oshima\thanks{E-mail: kooshima@gunma-ct.ac.jp}    \\ \\
\sl National Institute of Technology, Gunma College,Maebashi 371-8530, Japan}
\date{}
\maketitle
\begin{abstract}
We propose a simple procedure to produce energy eigenstates of a Hamiltonian with discrete eigenvalues.   
We use ancilla qubits and quantum entanglement
to separate an energy eigenstate from the other energy eigenstates.    
We exhibit a few examples derived from the (1+1)-dimensional massless Schwinger model.
Our procedure in principle will be applicable for a Hamiltonian with a finite dimensional Hilbert space. 
Choosing an initial state properly, we can in principle produce any energy eigenstate of the Hamiltonian.
\end{abstract} 
\newpage
\section{ Introduction}
 In quantum physics and quantum chemistry, it is an important task to find energy eigenstates of a quantum system.
For a large quantum system, however, it will be desperately difficult to find energy eigenstates  by the exact diagonalization of the 
Hamiltonian.   Since the publication of the universal type of quantum computer by IBM at 2016, quantum systems
have been actively analyzed by quantum computers\cite{Martinez1} and quantum simulators\cite{Honda1,Honda2,Okuda}.  A ground state is especially important among 
the energy-eigenstates and has been investigated actively. To prepare a ground state of a quantum
system in quantum computer, there are two representative methods; the adiabatic quantum computation\cite{Nishimori, Farhi} and the variational quantum eigensolver(VQE)\cite{Yung,Peruzzo}.
We can easily prepare an approximate ground state by the adiabatic quantum computation,  but we need relatively long
computation time.   Moreover, the approximate ground state is a superposition of a true ground state and excited states
with tiny but non-negligible amplitudes\cite{Honda1,Oshima1,Oshima2}.    As for the VQE, we can find a ground state rather well
with a few iterations.  Recently, (1+1)-dimensional Schwinger model\cite{Schwinger} has been studied with the IBM eagle processor by VQE\cite{Roland}.

  In this paper we propose a procedure to produce energy eigenstate of a Hamiltonian not only the ground state by the classically emulated quantum simulation.   We can easily carry out the procedure only using the
Hamiltonian and the ancilla qubits\cite{Nielsen} .  Concatenating ancilla qubits, we can in principle produce energy eigenstate precisely.   We operate controlled Hamiltonian
between the quantum system and the ancilla qubits.   The entropy of quantum entanglement between the quantum system and the ancilla
qubits increases and a certain energy eigenstate and the other energy eigenstates are separated.    In the previous paper\cite{Oshima1,Oshima2}, we have started from an approximate
ground state prepared by the adiabatic quantum computation.    This time we start from a state that is properly selected.
Choosing one initial state properly, we can produce one energy eigenstate.   We do not use the adiabatic quantum computation.   Therefore, the computation time
is very short.    We simulate a one-qubit system, a two-qubits system and a three-qubits system
based on the (1+1)-dimensional massless Schwinger model.

\section{One-qubit system} 
We consider the following one-qubit Hamiltonian.
\begin{equation}
{\hat H}=X+JZ,
\end{equation}
where $X$ and $Z$ are Pauli matrices $X= \left(\begin{array}{cc} 0 & 1 \\
1& 0 \end{array} \right)$, $Z=\left(\begin{array}{cc} 1 & 0 \\
0& -1 \end{array} \right)$ and $J$ is a real parameter.  The matrix form of the Hamiltonian is
\begin{equation}   
{\hat H}= \left(\begin{array}{cc} J & 1 \\
1& -J \end{array} \right),
\end{equation}
and its eigenvalues $\epsilon_{0},\epsilon_{1}$ and the corresponding eigenstates $|u_{0}\rangle, |u_{1}\rangle$ are
\begin{equation}
\epsilon_{0}=-\sqrt{J^{2}+1}, \qquad |u_{0}\rangle={1 \over \sqrt{2(J^{2}+1+J\sqrt{J^{2}+1})}} \left(\begin{array}{c} 1  \\
-\sqrt{J^{2}+1}-J \end{array} \right),
\end{equation}
\begin{equation}
\epsilon_{1}=\sqrt{J^{2}+1}, \qquad |u_{1}\rangle={1 \over \sqrt{2(J^{2}+1+J\sqrt{J^{2}+1})}} \left(\begin{array}{c} \sqrt{J^{2}+1}+J  \\
1 \end{array} \right).
\end{equation}
The expectation values of $Z$ are $\langle u_{1}|Z|u_{1}\rangle=-\langle u_{0}|Z|u_{0}\rangle={J^{2}+J\sqrt{J^{2}+1} \over J^{2}+1+J\sqrt{J^{2}+1}}$.

Using the Hamiltonian ${\hat H}$ and ancilla qubits, we produce the two eigenstates by the classically emulated quantum simulation.
First, we start from the initial state $|\psi_{0}\rangle=|0\rangle= \left(\begin{array}{c} 1 \\ 0 \end{array} \right)$.   We apply the first twirling operation in Fig.1 on $|\psi_{0}\rangle$; the quantum system time develops from $|\psi_{0}\rangle$ to $|\psi_{1}\rangle$ by the Hamiltonian controlled by the first ancilla qubit during a definite time period $\tau_{0}$.  This time period $\tau_{0}$
is prescribed by $\tau_{0}={\pi \over 2E_{0}}$, where $E_{0}=\langle \psi_{0}|{\hat H}|\psi_{0} \rangle$.   Next, we compute the expectation
value $E_{1}=\langle \psi_{1}|{\hat H}|\psi_{1}\rangle$.    We apply the second twirling operation on  $|\psi_{1}\rangle$
controlled by the second ancilla qubits.    This time the time period is  $\tau_{1}={\pi \over 2E_{1}}$.    We repeat this process for adequate
times.    After some rounds of the twirling operations the toatal state is $|\Psi\rangle=|\psi_{k}\rangle|0\rangle|0\rangle\cdots|0\rangle+|\phi_{k}\rangle||1\rangle\rangle$, where $||1\rangle\rangle$ means that at least one of the ancilla qubits is in the state $|1\rangle$.   We call the state
 $|\psi_{k}\rangle|0\rangle|0\rangle\cdots|0\rangle$ active state.  In Table I, we show our simulation result by IBM qiskit qasm-simulator for $J=1$.   
We have used the second order Suzuki-Trotter formula\cite{Trotter,Suzuki}.
We see that each eigenstate is produced depending on the initial state $|\psi_{0}\rangle$.     In addition to the energy we have measured $Z$ which does not commutes with the
Hamiltonian ${\hat H}$ and gives a character of goodness of the produced state.\\
\\
\begin{center}
\begin{tabular}{|c|c|c|c|c|c|c|c|} \hline
 $|\psi_{0}\rangle=|0\rangle$   &   0  &   1  & 2 & 3 & 4 & 5 & theoretical\\ \hline
$\langle{Z}\rangle$&1.0000000 &0.7972744
 &0.7022552 & 0.7072281& 0.7073355& 0.7073265&0.707107\\ \hline
$\langle{{\hat H}}\rangle$ &0.9999628 &1.3596955
 &1.4143788 & 1.4145616& 1.4145079&1.4144442 &1.414214\\ \hline
active state   & $10^{7}$  &  7813952  & 7656759 & 7655162  & 7656563 & 7657168&- \\ \hline \hline
 $|\psi_{0}\rangle=|1\rangle$   &   0  &   1  & 2 & 3 & 4 & 5 & theoretical\\ \hline
$\langle{Z}\rangle$ &1.0000000 &-0.7968211
 &-0.7020439 & -0.7069365& -0.7067735& -0.7072803&-0.707107\\ \hline
$\langle{{\hat H}}\rangle$ &-1.0001038 &-1.3594381
 &-1.4140652 & -1.4134110& -1.4141577&-1.4146414&-1.414214\\ \hline
active state   & $10^{7}$  &  6894003  & 6882968 & 6880064  & 6881986 & 6878468 &-\\ \hline
\end{tabular}
\end{center}
{} \quad \\
Table I.\qquad An example of average values of ${Z}, {\hat H}$ and the number of active states for each round of the twirling
operation for J=1.   The "active state" means the number of states that all the ancilla qubits are measured to be in the state $|0\rangle$.
The number $0$ means only measurements of  ${Z}$ and ${\hat H}$ have been done without twirling operations.
The numbers $1,2,\cdots,5$ indicate the number of times that we have acted the twirling operation.  We have used $n=10^{7}$ shots.
We exhibit the results for the initial state $|\psi_{0}\rangle=|0\rangle$ and for the initial state $|\psi_{0}\rangle=|1\rangle$.\\
\\
In the following, we explain why an energy eigenstate is produced.   A quantum state $|\psi\rangle$ is expanded as $|\psi\rangle=\alpha|u_{i}\rangle+\Sigma_{j \ne i}\beta_{j}|u_{j}\rangle$.   If we know the enegy $\epsilon_{i}$, by applying the corresponding twirling operation controlled by the ancilla qubit, the total state is changed as 
$|\psi\rangle{|0\rangle+|1\rangle \over \sqrt{2}} \rightarrow \alpha|u_{i}\rangle{|0\rangle+|1\rangle \over \sqrt{2}}+\Sigma_{j\ne i}\beta_{j}{|0\rangle+e^{i\theta_{j}}|1\rangle \over \sqrt{2}}$, where $\theta_{j}$ is an angle derived from the energy difference $\epsilon_{j}-\epsilon_{i}$.   By the Hadamard transformation on the ancilla qubit
we have the following total state $\alpha_{i}|u_{i}\rangle|0\rangle+\Sigma_{j \ne i}\beta_{j}|u_{j}\rangle{e}^{i{\theta_{j} \over 2}}(\cos{\theta_{j} \over 2}|0\rangle-i\sin{\theta_{j} \over 2}|1\rangle)$.    If we exclude the ancilla qubit in the state $|1\rangle$ the amplitude of the state $|u_{i}\rangle$ is relatively amplified.
Even when we only know an approximate value of $\epsilon_{i}$, the above situation will hold under the condition that the angle $\theta_{i}$ is small.  

\section{Two-qubits system}
In this section we introduce the following two-qubits Hamiltonian based on the (1+1)-dimensional massless Schwinger model\cite{Honda1,Okuda,Oshima2},
which is a (1+1)-dimensional Quantum Electrodynamics(QED),
\begin{equation}
 {\hat H}={1 \over 2}GZ_{0}+{1 \over 2}w(X_{0}X_{1}+Y_{0}Y_{1}),
\end{equation}
where $Y_{j}= \left(\begin{array}{cc} 0 & -i \\
i & 0 \end{array} \right)$ and $j=0,1$ means $j$-th qubit.
We have two parameters $G={1 \over 2}g^{2}a$ and $w={1 \over 2a}$, where $g$ is a coupling constant between an electron and  the Electric Field and $a$ is a
lattice spacing; an electron and a positron are created at $x=0$ and $x=a$.    We consider the following dimensionless Hamiltonian
\begin{equation}
{\hat H}={1 \over 2}(X_{0}X_{1}+Y_{0}Y_{1})+JZ_{0}, \quad J={g^{2}a^{2}} \ge 0.
\end{equation}
   In the matrix form the Hamiltonian is 
\begin{eqnarray}{\hat H}= \left(\begin{array}{cccc}
          J & 0 & 0 & 0 \\
          0     & J & 1 & 0 \\
          0    & 1 & -J & 0 \\
          0    &  0 & 0 & -J
\end{array} \right), 
\end{eqnarray}
and its eigenvalues $\epsilon_{0},\epsilon_{1},\epsilon_{2},\epsilon_{3} $ and the corresponding eigenstates $|u_{0}\rangle, |u_{1}\rangle,|u_{1}\rangle, |u_{2}\rangle$ are
\begin{equation}
\epsilon_{0}=-\sqrt{J^{2}+1}, \qquad |u_{0}\rangle={1 \over \sqrt{2(J^{2}+1+J\sqrt{J^{2}+1})}} \left(\begin{array}{c} 0\\ 1  \\
-\sqrt{J^{2}+1}-J \\ 0 \end{array} \right),
\end{equation}
\begin{equation}
\epsilon_{1}=-J, \qquad |u_{1}\rangle=\left(\begin{array}{c} 1\\ 0  \\
0\\ 0 \end{array} \right)=|0\rangle|0\rangle,
\end{equation}
\begin{equation} \epsilon_{2}=J, \qquad |u_{1}\rangle=\left(\begin{array}{c} 0\\ 0  \\
0\\ 1 \end{array} \right)=|1\rangle|1\rangle,
\end{equation}
\begin{equation}
\epsilon_{3}=\sqrt{J^{2}+1}, \qquad |u_{3}\rangle={1 \over \sqrt{2(J^{2}+1+J\sqrt{J^{2}+1})}} \left(\begin{array}{c} 0 \\\sqrt{J^{2}+1}+J  \\
1 \\ 0 \end{array} \right).
\end{equation}
In the same way as the one-qubit case, using the Hamiltonian ${\hat H}$ and ancilla qubits, we produce the four eigenstates by the classically emulated quantum simulation.
The quantum circuit for the two-qubits system is shown in Fig.2.   We use two ancilla qubits for one twirling operation.
As the initial state of the simulation we choose four states $|0\rangle_{0}|0\rangle_{1}$, $|0\rangle_{0}|1\rangle_{1}$,$|1\rangle_{0}|0\rangle_{1}$ and $|1\rangle_{0}|1\rangle_{1}$. 
In terms of the (1+1)-dimensional Schwinger model, the states $|0\rangle_{0}|1\rangle_{1}$ and $|1\rangle_{0}|0\rangle_{1}$ correspond to the state that no electrons nor
no positrons exist and the state that a pair of an electron and a positron bounded by the Electric Field is created, respectively.  The states $|0\rangle_{0}|0\rangle_{1}$ and $|1\rangle_{0}|1\rangle_{1}$ correspond to the state that one positron exists and the state that one electron exists, respectively.    The latter two states undergo no changes
under the charge preserving Hamiltonian.   In Table II, we show our simulation result for $J=1$.   We see that
each eigenstate is produced depending on the initial state $|\psi_{0}\rangle$.     In addition to the energy we have measured $Z_{0}$ which does not commutes with the
Hamiltonian ${\hat H}$.\\
\\
\begin{center}
\begin{tabular}{|c|c|c|c|c|c|c|c|} \hline
 $|0\rangle_{0}|1\rangle_{1}$   &   0  &   1  & 2 & 3 & 4 & 5 & theoretical\\ \hline
$\langle{Z_{0}}\rangle$&1.0000000 &0.6863952
 &0.7073591 & 0.7068051& 0.7073355& 0.7068394&0.707107\\ \hline
$\langle{{\hat H}}\rangle$ &0.9997964 &1.4084719
 &1.4146696 & 1.4143585& 1.4139457&1.4146494 &1.414214\\ \hline
active state   & $10^{7}$  &  6891253  & 6876931 & 6877308 & 6879307 & 6878882 &- \\ \hline \hline
 $|1\rangle_{0}|0\rangle_{1}$   &   0  &   1  & 2 & 3 & 4 & 5 & theoretical\\ \hline
$\langle{Z_{0}}\rangle$ &1.0000000 &-0.6863618
 &-0.7070429 & -0.7069414& -0.7069884& -0.7073392 &-0.707107\\ \hline
$\langle{{\hat H}}\rangle$ &-0.999909 &-1.40843695
 &-1.4142024 & -1.4138955& -1.4143327&-1.4151875 &-1.414214\\ \hline
active state   & $10^{7}$  &  6893632  & 6880351 & 6881961  & 6878923& 6875563 &-\\ \hline
\end{tabular}
\end{center}
{} \quad \\
Table II \qquad An example of average values of ${Z_{0}}, {\hat H}$ and the number of active states for each round of the twirling
operation for J=1.   We have used $n=10^{7}$ shots.
We exhibit the results for the initial state $|\psi_{0}\rangle=|0\rangle_{0}|1\rangle_{1}$ and for the initial state $|\psi_{0}\rangle=|1\rangle_{0}|0\rangle_{0}$.
The trivial cases $|\psi_{0}\rangle=|0\rangle_{0}|0\rangle_{1}$ and  $|\psi_{0}\rangle=|1\rangle_{0}|1\rangle_{0}$ are omitted.

\section{Three-qubits system}
In this section we consider the three-qubits Hamiltonian  based on the $(1+1)$-dimensional massless Schwinger model\cite{Honda1,Okuda,Oshima2}.  
We now have the following Hamiltonian \\
\begin{equation}
 {\hat H}={1 \over 2}G(Z_{0}+Z_{0}Z_{1})+{1 \over 2}w(X_{0}X_{1}+X_{1}X_{2}+Y_{0}Y_{1}+Y_{1}Y_{2}).
\end{equation}
We consider the corresponding dimensionless Hamiltonian\\
\begin{equation}
 {\hat H}={1 \over 2}(X_{0}X_{1}+X_{1}X_{2}+Y_{0}Y_{1}+Y_{1}Y_{2})+{1 \over 2}J(Z_{0}+Z_{0}Z_{1}).
\end{equation}

This Hamiltonian has 8 energy eigenstates.    We label them as $|E_{0}\rangle, |E_{1}\rangle, \cdots, |E_{7}\rangle$, where $|E_{0}\rangle$ is the ground state and $|E_{i}\rangle$ is the 
$i$-th excited state.   By simple algebraic calculation we can find these states.     We see that
\begin{eqnarray}
 |E_{7}\rangle&=&{1 \over \sqrt{4(1+6J^{2}+8J^{4})+8\sqrt{2}J(1+2J^{2})^{3 \over 2}} }((1+4J^{2}+2\sqrt{2}J\sqrt{1+2J^{2}})|0\rangle_{0}|0 \rangle_{1}|1\rangle_{2} \nonumber \\
&+&(2J+\sqrt{2}\sqrt{1+2J^{2}})|0\rangle_{0}|1 \rangle_{1}|0\rangle_{2}+|1\rangle_{0}|0 \rangle_{1}|0\rangle_{2}),
\end{eqnarray}
\begin{equation}
  |E_{6}\rangle=|0\rangle_{0}|0 \rangle_{1}|0\rangle_{2},
\end{equation}
\begin{eqnarray}
|E_{5}\rangle={1 \over \sqrt{4+2J^{2}-2J\sqrt{2+J^{2}}}}(|0\rangle_{0}|1\rangle_{1}|1\rangle_{2}+(-J+\sqrt{2+J^{2}})|1\rangle_{0}|0\rangle_{1}|1\rangle_{2}+|1\rangle_{0}|1\rangle_{1}0\rangle_{2}), 
\end{eqnarray} 
\begin{equation}
|E_{4}\rangle={1 \over \sqrt{2+4J^{2}}}(-|0\rangle_{0}|0\rangle_{1}|1\rangle_{2}+ 2J|0\rangle_{0}|1\rangle_{1}|0\rangle_{2}+|1\rangle_{0}|0\rangle_{1}|0\rangle_{2}),
\end{equation}
\begin{equation}
|E_{3}\rangle={1 \over \sqrt{2}}(-|0\rangle_{0}|1\rangle_{1}|1\rangle_{2}+|1\rangle_{0}|1\rangle_{1}|0\rangle_{2}),
\end{equation}
\begin{equation}
  |E_{2}\rangle=|1\rangle_{0}|1 \rangle_{1}|1\rangle_{2},
\end{equation}
\begin{eqnarray} 
|E_{1}\rangle &=&{1 \over \sqrt{4(1+6J^{2}+8J^{4})+8\sqrt{2}J(1+2J^{2})^{3 \over 2}} }((1+4J^{2}-2\sqrt{2}J\sqrt{1+2J^{2}})|0\rangle_{0}|0 \rangle_{1}|1\rangle_{2} \nonumber \\
&+&(2J-\sqrt{2}\sqrt{1+2J^{2}})|0\rangle_{0}|1 \rangle_{1}|0\rangle_{2}+|1\rangle_{0}|0 \rangle_{1}|0\rangle_{2}),
\end{eqnarray}
\begin{equation}
|E_{0}\rangle={1 \over \sqrt{4+2J^{2}+2J\sqrt{2+J^{2}}}}(|0\rangle_{0}|1\rangle_{1}|1\rangle_{2}-(J+\sqrt{2+J^{2}})|1\rangle_{0}|0\rangle_{1}|1\rangle_{2}+|1\rangle_{0}|1\rangle_{1}|0\rangle_{2}).
\end{equation}
The corresponding eigenvalues are $\epsilon_{7}=\sqrt{2}\sqrt{1+2J^{2}}, \epsilon_{6}=2J, \epsilon_{5}=\sqrt{2+J^{2}}-J, \epsilon_{4}=\epsilon_{3}=\epsilon_{2}=0, \epsilon_{1}=-\sqrt{2}\sqrt{1+2J^{2}}$  and $\epsilon_{0}=-(J+\sqrt{2+J^{2}}).$    Except for the energy,  we measure the physical quantity ${\bar Z}={1 \over 3}(Z_{0}-Z_{1}+Z_{2})$ 
which does not commute with the Hamiltonian.    The expectation values of ${\bar Z}$ for the above energy eigenstates for $J=1$ are
\begin{eqnarray}
\langle E_{7}| {\bar Z}|E_{7} \rangle=-0.11111, \quad \langle E_{6}| {\bar Z}|E_{6} \rangle =0.33333, \quad \langle E_{5}| {\bar Z}|E_{5} \rangle =0.05157, \quad \langle E_{4}| {\bar Z}|E_{4} \rangle =-0.33333 \nonumber \\
\langle E_{3}| {\bar Z}|E_{3} \rangle=0.33333, \quad \langle E_{2}| {\bar Z}|E_{2} \rangle =0.55556, \quad \langle E_{1}| {\bar Z}|E_{1} \rangle =-0.11117, \quad \langle E_{0}| {\bar Z}|E_{0} \rangle =-0.71823.
\end{eqnarray}
\\
In the same way as in the two-qubits case, starting from an adequate initial state we can produce a corresponding energy eigenstate.
For a three-qubits Hamiltonian we need three ancilla bits for one twirling operation(see Fig.3).  

Starting from the initial state $|0\rangle_{0}|0\rangle_{1}|1\rangle_{2}$ we can produce the state $|E_{7}\rangle$.    We show a simulation result in Table III.
The state $|E_{7}\rangle$ is a superposition of the initial state $|0\rangle_{0}|0\rangle_{1}|1\rangle_{2}$ and other states.
\newpage
\begin{center}
\begin{tabular}{|c|c|c|c|c|c|c|} \hline
 $|0\rangle_{0}|0\rangle_{1}|1 \rangle_{2}$   &   0          &   1         & 2     &  3   & 4 &theoretical \\ \hline
$\langle{{\bar Z}}\rangle$                      &-0.33333   &-0.111612& -0.11114&  -0.11124& -0.11109 &-0.11111\\ \hline
$\langle{{\hat H}}\rangle$                     &2.00040     &2.44939  & 2.45003 & 2.44933&2.44949& 2.44895 \\ \hline
active states                                    &$10^{7} $    &  4083826& 4080271 & 4083011 & 4082364  & - \\ \hline
\end{tabular}
\end{center}
{} \qquad \\
Table III.    Pruduction of $|E_{7}\rangle$.   An example of average values of ${\bar Z}, {\hat H}$ and the number of active states for each round of the twirling
operation for J=1 is shown.   We have used $n=10^{7}$ shots.   The initial state is $|\psi_{0}\rangle=|0\rangle_{0}|0\rangle_{1}|1\rangle_{2}$.\\

Starting from the initila state $|0\rangle_{0}|0\rangle_{1}|0\rangle_{2}$ we can produce the state $|E_{6}\rangle$.    We show a simulation result in Table IV.
Since $|0\rangle_{0}|0\rangle_{1}|0\rangle_{2}$ is $|E_{6}\rangle$ itself,  we have a trivial result.\\

\begin{center}
\begin{tabular}{|c|c|c|c|} \hline
 $|0\rangle_{0}|0\rangle_{1}|0 \rangle_{2}$   &   0  &   1  & theoretical \\ \hline
$\langle{{\bar Z}}\rangle$& 0.33333            &0.33333  &0.33333\\ \hline
$\langle{{\hat H}}\rangle$& 2.00045           &2.00037 &2 \\ \hline
active states  &$10^{7} $  &   9999991    & - \\ \hline
\end{tabular}
\end{center}
{} \qquad \\
Table IV.   Certification of $|E_{6}\rangle$.    Under the twirling operation no meaningful changes occur.\\

Starting from the initila state $|0\rangle_{0}|1\rangle_{1}|1\rangle_{2}$ we intend to produce the state $|E_{5}\rangle$.   In this case, since the expectation value of the Hamiltonian is almost zero,  we are in a delicate situation.   We cannot use the time $\tau_{0}={\pi \over 2E_{0}}$.  We explore a suitable value of $\tau_{0}$ by experience. 
We show a simulation result in Table V that we have replaced the value of $E_{0}$ by $E=0.5$ artificially.   We see that the state $|E_{5}\rangle$ seems to be produced.    Since the number of active state is very small, the expectation values of ${\bar Z}$ and ${\hat H}$ correspondingly sift from the theoretical values.\\

\begin{center}
\begin{tabular}{|c|c|c|c|c|c|c|} \hline
 $|0\rangle_{0}|1\rangle_{1}|1 \rangle_{2}$   &   0          &   1(E=0.5)         & 2     &  3   & 4 &theoretical \\ \hline
$\langle{{\bar Z}}\rangle$                      &0.33333   &-0.17105& 0.05702&  0.0548& 0.04729 &0.05157\\ \hline
$\langle{{\hat H}}\rangle$                     &0.00034   &0.60887  & 0.72180 & 0.72684 &0.72715& 0.73205 \\ \hline
active states                                    &$10^{7} $    & 25930& 1431771 & 14142 & 14070  & - \\ \hline
\end{tabular}
\end{center}
{} \qquad \\
Table V.    Production of $|E_{5}\rangle$.   For the initial state, since the expectation value of the Hamiltonian is almost zero, in the first 
twirling operation we set $E=0.5$ by the hands. \\
\\
\\
We show in Table VI another simulation result for the initial state  $|0\rangle_{0}|1\rangle_{1}|1\rangle_{2}$.    In this case the ground state $|E_{0}\rangle$ is produced,
which will be attributed that the two state $|E_{5}\rangle$ and $|E_{0}\rangle$  are akin in the sense the two states are linear combinations of  $|0\rangle_{0}|1\rangle_{1}|1 \rangle_{2}$, 
$|1\rangle_{0}|0\rangle_{1}|1 \rangle_{2}$ and $|1\rangle_{0}|1\rangle_{1}|0 \rangle_{2}$ . \\

\begin{center}
\begin{tabular}{|c|c|c|c|c|c|c|} \hline
 $|0\rangle_{0}|1\rangle_{1}|1 \rangle_{2}$   &   0          &   1(E=0.4)         & 2     &  3   & 4 &theoretical \\ \hline
$\langle{{\bar Z}}\rangle$                      &0.33333   &-0.25223& -0.79754&  -0.72597& -0.72016 &0.05157\\ \hline
$\langle{{\hat H}}\rangle$                     &0.00030   &-2.19846  & -2.71408 & -2.73546 &-2.72963& 0.73205 \\ \hline
active states                                    &$10^{7} $    & 843541& 311532 & 309589 & 310191  & - \\ \hline
\end{tabular}
\end{center}
{} \qquad \\
Table VI.    Failure to produce $|E_{5}\rangle$.      In the first twirling operation we set $E=0.4$ by the hands.    In this case the ground state $|E_{0}\rangle$ seems to be produced.    For the value $E=0.4$, the state $|E_{5}\rangle$ is not produced. \\

Next, we will treat the three zero energy states.   For the non-zero energy states,  we have adopted the twirling time period $\tau={\pi \over 2E}$ and have used the relation $ie^{-i{\pi \over 2}}=1$ .   If $E=\langle E|{\hat H}|E\rangle=0$ as in the case $|E_{5}\rangle$,  we cannot directly adopt the relation $\tau={\pi \over 2E}$. 
Note that, if ${\hat H}|E\rangle=0|E\rangle$, we have $e^{-i\hat H\tau}|E\rangle=|E\rangle$ for any $\tau$. 
Under this situation, to produce the zero energy states we adopt the twirling time period $\tau={2\pi \over E}$ for some artificial non-zero value $E$ and  use the simple relation $e^{-2\pi{i}}=1$ not the relation $ie^{-i{\pi \over 2}}=1$.

We produce the state $|E_{4}\rangle$.   We examine the initial state $|0\rangle_{0}|1\rangle_{1}|0\rangle_{2}$.   The expectation value of the Hamiltonian for this state is 0.
In the following we see that this value 0 is not violated by the twirling operations.    We perform the twirling operations four times by setting the value of $E$ as $E=-0.2$ 
or $E=0.2$ at will.   We show a result in Table VII.     The value ${\bar Z}$ varies violently  and it seems  that the state $|E_{4}\rangle$ may not be produced by the four times 
twirling operations.

\begin{center}
\begin{tabular}{|c|c|c|c|c|c|c|} \hline
 $|0\rangle_{0}|1\rangle_{1}|0 \rangle_{2}$   &   0  &   1($E=-0.2$)        & 2 ($E=0.2$)          & 3 ($E=0.2$)         & 4($E=-0.2$)            & theoretical \\ \hline
$\langle{{\bar Z}}\rangle$&1.00000             &0.51226 &0.86966 & 0.57704& 0.56051&0.55556\\ \hline
$\langle{{\hat H}}\rangle$ &0.00014 & 0.00062   &-0.00015&0.00087 & 0.00020 &0 \\ \hline
active state                 & $10^{7}$                 &  8110010 & 7288765 & 6929521  & 6669713 &   - \\ \hline
\end{tabular}
\end{center}
{} \quad \\
Table VII  Attempt to produce $|E_{4}\rangle$.   We cannot be convinced that the state $|E_{4}\rangle$ is produced.\\
\\
This time, we set $E=-1$, $E=2$, $E=-3$ and $E=4$ in the four twirling operations.   We show a result in Table VIII.   We see that the state $|E_{4}\rangle$ is produced almost exactly.\\
\\
\begin{center}
\begin{tabular}{|c|c|c|c|c|c|c|} \hline
 $|0\rangle_{0}|1\rangle_{1}|0 \rangle_{2}$   &   0  &   1($E=-1$)        & 2 ($E=2$)          & 3 ($E=3$)         & 4($E=-4$)            & theoretical \\ \hline
$\langle{{\bar Z}}\rangle$&1.00000             &0.57776 &0.56048 & 0.55591& 0.55560 &0.55556\\ \hline
$\langle{{Z_{0}}}\rangle$&1.00000             &0.68331 &0.67084 & 0.66684& 0.66678 &0.66667\\ \hline
$\langle{{Z_{1}}}\rangle$&-1.00000             &-0.36665 &-0.34071 & -0.33387& -0.33394 &-0.33333\\ \hline
$\langle{{Z_{2}}}\rangle$&1.00000             &0.68333 &0.66988 & 0.66703& 0.66615 &0.66667\\ \hline
$\langle{{\hat H}}\rangle$ &0.00007 & -0.00003   &0.00135&-0.00018 & 0.00052 &0 \\ \hline
active state                 & $10^{7}$                 &  6671011 & 6666402 & 6665359  & 666450 &   - \\ \hline
\end{tabular}
\end{center}
{} \quad \\
Table VIII   Production of $|E_{4}\rangle$.  In addition to ${\bar Z}$ and ${\hat H}$, $Z_{0}$, $Z_{1}$ and $Z_{2}$ are measured. \\

Starting from the initial state $|1\rangle_{0}|1\rangle_{1}|0\rangle_{2}$, we produce the state $|E_{3}\rangle$.   Our simulation result is in Table IX.    Since the expectation
value of ${\hat H}$ is almost zero for $|1\rangle_{0}|1\rangle_{1}|0\rangle_{2}$ we simulate by setting $E=-0.2$ for the first twirling operation.   We also take $E=0.2$ for the third
twirling operation and take $E=-0.2$ for the fourth twirling operation.    We see that the state $|E_{3}\rangle$ is produced with high accuracy.

\begin{center}
\begin{tabular}{|c|c|c|c|c|c|c|} \hline
 $|1\rangle_{0}|1\rangle_{1}|0 \rangle_{2}$   &   0  &   1($E=-0.2$)        & 2           & 3 ($E=0.2$)         & 4($E=-0.2$)            & theoretical \\ \hline
$\langle{{\bar Z}}\rangle$&0.33333            &0.29169 &0.33292 & 0.33332& 0.33333 &0.33333\\ \hline
$\langle{{Z_{0}}}\rangle$&-1.00000             &0.10780 &0.02292 & -0.01136& 0.00524 &0\\ \hline
$\langle{{Z_{1}}}\rangle$&-1.00000             &-0.93754 &-0.99938 & -0.99998& -0.99999 &-1\\ \hline
$\langle{{Z_{2}}}\rangle$&1.00000             &-0.17026 &-0.02354 & 0.01134& -0.00524 &0\\ \hline
$\langle{{\hat H}}\rangle$ &0.00003 & 0.06916   &0.00008&0.00005 & 0.00001 &0 \\ \hline
active state                 & $10^{7}$                 &  5588131 & 5002241 & 5001886  & 5001862 &   - \\ \hline
\end{tabular}
\end{center}
{} \quad \\
Table IX.  Production of $|E_{3}\rangle$. \quad In the second twirling operation the non-zero value $E_{1}=0.06916$ has been used.\\

We start from $|1\rangle_{0}|1\rangle_{1}|1\rangle_{2}$ which is $|E_{2}\rangle$ itself.   This time the expectation value of the Hamiltonian is almost 0 and we cannot set the twirling time period 
as $\tau_{0}={2\pi \over E}$.   Therefore we perform
the first twirling operation by setting $E=-0.2$, which is a suitable value other than 0.   We show a simulation result in the Table X.    We see that the state $|1\rangle_{0}|1\rangle_{1}|1\rangle_{2}$ is
almost not changed by the twirling operation and is
a zero energy eigenstate of ${\hat H}$. \\

\begin{center}
\begin{tabular}{|c|c|c|c|} \hline
 $|1\rangle_{0}|1\rangle_{1}|1 \rangle_{2}$   &   0  &   1($E=-0.2$)  & theoretical \\ \hline
$\langle{{\bar Z}}\rangle$&-0.33333            &-0.33333  &-0.33333\\ \hline
$\langle{{\hat H}}\rangle$&-0.00011           &-0.00016 &0 \\ \hline
active states  &$10^{7} $  &   9999999    & - \\ \hline
\end{tabular}
\end{center}
{} \qquad \\
Table X.   Certification of $|E_{2}\rangle$.  We have assumed the artificial value $E=-0.2$ in the first twirling operation.\\

Starting from the initial state $|1\rangle_{0}|0\rangle_{1}|0\rangle_{2}$ we can produce the first exited state $|E_{1}\rangle$.    We show a simulation result in Table.XI.\\

\begin{center}
\begin{tabular}{|c|c|c|c|c|c|c|} \hline
 $|1\rangle_{0}|0\rangle_{1}|0 \rangle_{2}$   &   0          &   1         & 2     &  3   & 4 &theoretical \\ \hline
$\langle{{\bar Z}}\rangle$                      &-0.33333   &-0.11613& -0.11092&  -0.11123& -0.11107 &-0.11117\\ \hline
$\langle{{\hat H}}\rangle$                     &-2.00046     &-2.44702  & -2.454908 & -2.44884&2.44984& -2.44949 \\ \hline
active states                                    &$10^{7} $    &  4083447& 4083541 & 4083922 & 4082517  & - \\ \hline
\end{tabular}
\end{center}
{} \qquad \\
Table XI.    Pruduction of $|E_{1}\rangle$.    The time period $\tau={\pi \over 2E}$ and the relation $ie^{-i{\pi \over 2}}=1$ are used. \\

Starting from the initila state $|1\rangle_{0}|0\rangle_{1}|1\rangle_{2}$ we can produce the ground state $|E_{0}\rangle$.    We show a simulation result in Table XII.\\

\begin{center}
\begin{tabular}{|c|c|c|c|c|c|c|} \hline
 $|1\rangle_{0}|0\rangle_{1}|1 \rangle_{2}$   &   0          &   1         & 2     &  3   & 4 &theoretical \\ \hline
$\langle{{\bar Z}}\rangle$                      &-1.00000   &-0.520733& -0.70065&  -0.71411& -0.71856 &-0.71823\\ \hline
$\langle{{\hat H}}\rangle$                     &-1.99963     &-2.63677  & -2.73484 & -2.73240&-2.73279& -2.73205 \\ \hline
active states                                    &$10^{7} $    &  874150& 834283 & 834484 & 835523  & - \\ \hline
\end{tabular}
\end{center}
{} \qquad \\
Table XII.    Production of $|E_{0}\rangle$.     The adiabatic quantum computation is not used to produce the ground state $|E_{0}\rangle$.\\

To produce  a ground state we can use the adiabatic quantum computation.  We adopt the initial trivial Hamiltonian as ${\hat H}_{0}=Z_{0}-Z_{1}+Z_{2}$,
and we start from the initial ground state $|1\rangle_{0}|0\rangle_{1}|1\rangle_{2}$.   After the adiabatic quantum computation we have almost ground state, 
that is a superposition of a ground state with excited states with small amplitudes.     We show a simulation result in Table XIII.
After the three or four twirling operations the results in the Table XII and in the Table XIII  are almost indistinguishable. \\
\\

\begin{center}
\begin{tabular}{|c|c|c|c|c|c|c|} \hline
 $|1\rangle_{0}|0\rangle_{1}|1 \rangle_{2}$   &   adiabatic          &   1         & 2     &  3   & 4 &theoretical \\ \hline
$\langle{{\bar Z}}\rangle$                      &-0.71256   &-0.71863& -0.71817&  -0.71824& -0.71859 &-0.71823\\ \hline
$\langle{{\hat H}}\rangle$                     &-2.73169    &-2.73227  & -2.73221 & -2.73246&-2.73281& -2.73205 \\ \hline
active states                                    &$10^{7} $    &  9998684& 9998546 & 9998621 & 9998620  & - \\ \hline
\end{tabular}
\end{center}
{} \qquad \\
Table XIII.    Production of $|E_{0}\rangle$ by the adiabatic quantum computation.   Before the twirling operations we have used the adiabatic quantum computation. \\

\section{Summary and discussions}
 We have proposed a procedure to produce energy eigenstate of a quantum Hamiltonian without algebraic calculation.   We have performed classically emulated
quantum simulation for the one-qubit system , the two-qubits system and the three-qubits system based on the (1+1)-dimensional massless Schwinger model.  
Choosing an initial states properly, we have produced the corresponding energy eigenstate.   We have concatenated the ancilla qubits to rectify one energy eigenstate.
Our procedure is in principle applicable to a Hamiltonian represented by the Pauli matrices.    Our procedure will work well for a Hamiltonian
with discrete eigenvalues.    An $n$-qubits Hamiltonian has $2^{n}$ energy-eigenstates.     In our procedure, choosing $2^{n}$ orthogonal initial
states properly, the $2^{n}$ energy eigenstates will be produced.   Without using the adiabatic quantum computation, we can produce
a ground state precisely by the twirling operation.   Since, we do not use the adiabatic quantum computation each computation time is very short.
As for the three-qubits case the four states $|E_{7}\rangle$, $|E_{6} \rangle$, $|E_{1} \rangle$ and $|E_{0} \rangle$ are produced or certified straightforwardly 
in our procedure.    The rest four states  $|E_{5}\rangle$, $|E_{4} \rangle$, $|E_{3} \rangle$ and $|E_{2} \rangle$ are produced or certified by setting some
artificial values of twirling period $\tau$.

\newpage
Figure captions.\\
\\
Fig 1. \\
One-qubit quantum circuit.  The ancilla bits are initialized in the state $|0\rangle$.  The unitary operator $U(\tau)$, which we call the twirling operator, is $U(\tau)=ie^{-i{\tau}{\hat H}}$.
We can choose the initial state $|\psi_{0}\rangle$ properly.   We have chosen $|\psi_{0}\rangle=|0\rangle$ and $|\psi_{0}\rangle=|1\rangle$.
\\ \\
Fig 2. \\
Two-qubits quantum circuit.  The state $|\psi_{0}\rangle$ is a two-qubits initial state.   A pair of ancilla qubits is used for one twirling operation.
The figure shows three twirling operations.\\
\\ \\
Fig 3. \\
Three-qubits quantum circuit.  The state $|\psi_{i-1}\rangle$ is a three-qubits state.   Three ancilla qubits are used for the $i$-th twirling operation.
\\
\\ \\
\newpage
\begin{figure}[htbt]
 \includegraphics[keepaspectratio, scale=1.0] {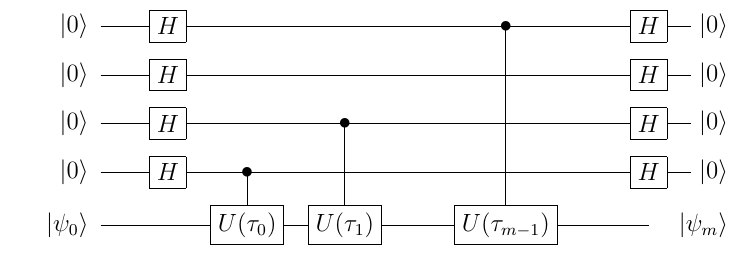}
\end{figure}
{} \qquad \qquad \qquad \qquad \qquad  \qquad \qquad \qquad \qquad  Fig.1\\
\\
\begin{figure}[htbt]
 \includegraphics[keepaspectratio, scale=0.8] {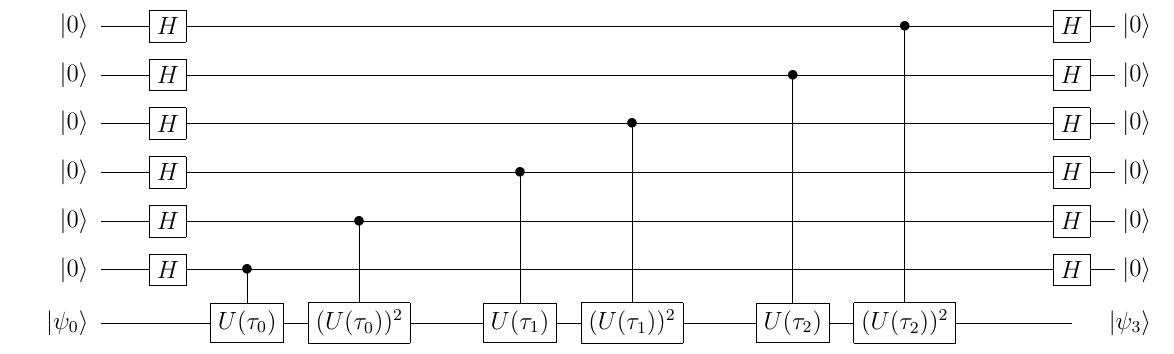}
\end{figure}
\\
{} \qquad \qquad \qquad \qquad \qquad  \qquad \qquad \qquad \qquad  Fig.2\\
\\
\begin{figure}[htbt]
 \includegraphics[keepaspectratio, scale=1] {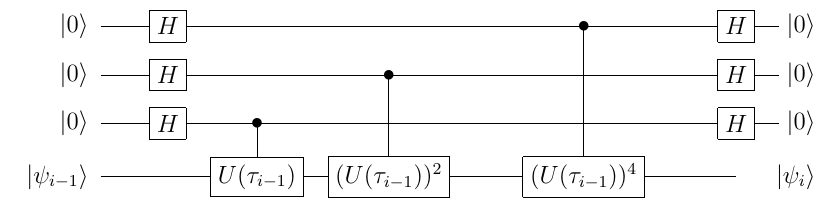}
\end{figure}
\\
{} \qquad \qquad \qquad \qquad \qquad  \qquad \qquad \qquad \qquad  Fig.3\\

\newpage
\begin{thebibliography}{99}
\bibitem{Martinez1}
E.A.Martinez, C.A. Muschik, P.Schindler, D.Nigg, A.Erhard, M.Heyl, P.Hauke, M.Dalmonte, T.Monz, P.Zoller and R.Blatt, {\bf 534}, 516 (2016).

\bibitem{Honda1}
B.Chakraborty, M.Honda, T.Izubuti, Y.Kikuchi and A.Tomiya, Phys. Rev. D{\bf 105}, 094503 (2022) .

\bibitem{Honda2}
M.Honda.E.Itou, Y.Kikuchi and Y.Tanizaki, PTEP{\bf 2022}, 033B01(2022).

\bibitem{Okuda}
 M.Honda, E.Itou, Y.Kikuchi, L.Nagano and T.Okuda, Phys. Rev. D{\bf 105}, 014504 (2022).

\bibitem{Nishimori}
T.Kadowaki and H.Nishimori, Phys.Rev.E{\bf 58}, 5355(1998).

\bibitem{Farhi}
       E.Farhi, J.Goldstone, S.Gutmann, J.Lapan, A.Lundgren and D.Preda, Science{\bf 292}, 472(2001).

\bibitem{Yung}
M.-H.Yung, J.Casanova, A.Mezzacapo, J.McClean, L.Lamata, A.Asupuru-Guzik and E.Solano, Sci.Per.{\bf 4},358982014).

\bibitem{Peruzzo}
A.Peruzzo, J.McClean,P.Shadbolt, M.-H.Yung, X.-Q.Zhou,P.J.Love, A.Asupuru-Guzik and J.L.${\rm O^{'}Brien}$, Nat.Comm.{\bf 5},4213(2014).

\bibitem{Oshima1}
K.Oshima, IET Quant. Comm, {\bf 3}, 214(2022).

\bibitem{Oshima2}
K.Oshima, arXiv:2308.15066 [quant-ph](2023), 

\bibitem{Schwinger}
{ J.S.Schwinger, Phys.Rev.{\bf 125}, 397(1962). }

\bibitem{Roland}
R. C. Farrell, M. Illa, A.N. Ciavarella and M.J. Savage, arXiv:2308.04481 [quant-ph](2023), 

\bibitem{Nielsen}
 M.A.Nielsen and I.L.Chuang, Quantum Computation and Quantum
    Information, Cambridge University Press, Cambridge, England, 2000. 

\bibitem{Trotter}
H.H.Trotter, Proc. Amer. Math. Soc. {\bf 10} ,545(1959).

\bibitem{Suzuki}
M.Suzuki, Comm. Math. Phys. {\bf 51}, 183 (1976). 


\end{thebibliography}
\end{document}